\documentclass[aps,prl,twocolumn,superscriptaddress,floatfix,amsmath]{revtex4}

\usepackage{graphicx}

\usepackage{amsmath,amssymb}

\begin{document}



\title{Exact results for curvature-driven coarsening in two dimensions}


\author{Jeferson J. Arenzon}
\affiliation{Instituto de F\'\i sica, Universidade Federal do 
Rio Grande do Sul, CP 15051, 91501-970 Porto Alegre RS, Brazil} 
\author{Alan J. Bray}
\affiliation{School of Physics and Astronomy, University of Manchester, 
Manchester M13 9PL, UK}
\author{Leticia F. Cugliandolo}
\affiliation{Universit\'e Pierre et Marie Curie -- Paris VI, LPTHE UMR 7589,
4 Place Jussieu,  75252 Paris Cedex 05, France}
\author{Alberto Sicilia}
\affiliation{Universit\'e Pierre et Marie Curie -- Paris VI, LPTHE UMR 7589,
4 Place Jussieu,  75252 Paris Cedex 05, France}


\date{\today}

\begin{abstract}
We consider the statistics of  the areas enclosed by domain boundaries
(`hulls')  during  the  curvature-driven  coarsening dynamics  of  a
two-dimensional  nonconserved scalar field  from a  disordered initial
state.  We show that the number of hulls per unit area that enclose an
area  greater than  $A$  has, for  large  time $t$,  the scaling  form
$N_h(A,t) = 2c/(A+\lambda t)$, demonstrating the validity of dynamical 
scaling in this system, where $c=1/8\pi\sqrt{3}$ is a universal constant.  
Domain areas (regions of aligned spins) have a similar 
distribution  up to very large values of $A/\lambda t$. Identical 
forms are  obtained for coarsening from a critical initial state, but 
with $c$ replaced by $c/2$.
\end{abstract}


\maketitle



Coarsening dynamics  has attracted enormous interest over  the last 40
years.  The classic  scenario concerns  a system  that  in equilibrium
exhibits a  phase transition from a  disordered high-temperature phase
to  an ordered  low-temperature phase with a broken symmetry of  the
high-temperature phase. The simplest example is, perhaps, the
Ising  ferromagnet. When  the  system is  cooled  rapidly through  the
transition  temperature, domains of  the two  ordered phases  form and
grow (`coarsen')  with time under  the influence of  the interfacial
surface tension, which acts as a driving force for the domain growth 
\cite{Lifshitz,AC,BrayReview}. 

While phase  transitions provide the traditional  arena for coarsening
dynamics,  there  are  many  other  examples,  including  soap  froths
\cite{tam},    breath   figures  \cite{Marcos-Martin}, 
granular media \cite{Aranson}, and interfacial
fluctuations \cite{Golubovic}.  A  common feature of
nearly all such coarsening systems  is that they are well described by
a   dynamical  scaling   phenomenology  in   which  there   is a single
characteristic length  scale,  $R(t)$,  which  grows  with  time.  If
dynamical scaling  holds, the  domain morphology is  statistically the
same at all times when all lengths  are measured in units of $R(t)$.
The   assumption  of   dynamical   scaling  also   makes  possible  the
determination  of  the  length  scale  $R(t)$ for  a  large  class  of
coarsening systems \cite{BrayReview,RutenbergBray}.  

Despite  the   success  of   the  scaling  hypothesis   in  describing
experimental and  simulation data, its  validity has only  been proved
for very simple models,  including the 
$1d$ Glauber-Ising model \cite{Amar-Family} and the nonconserved $O(n)$ model 
in the limit $n \to \infty$ \cite{Coniglio-Zannetti}. Another noteworthy 
exact result  is the Lifshitz-Slyozov derivation  of the  domain-size  
distribution for  a conserved  scalar field in  the limit  where the 
minority  phase occupies  a vanishingly small volume fraction 
\cite{Lifshitz-Slyozov}. The only other exact results, to our  
knowledge, for domain-size distributions in coarsening dynamics are 
for the zero-temperature Glauber-Potts \cite{Derrida-Zeitak} 
and time-dependent Ginzburg-Landau \cite{BDG} models in $1d$.  

In the  present work we obtain  some exact results  for the coarsening
dynamics   of  a   nonconserved  scalar   field  in   $d=2$,
demonstrating explicitly, {\em en passant}, the  validity  of  the  
scaling  hypothesis. To do this, we use  a continuum 
model in which the  velocity, $v$, of each element of a domain boundary  
is proportional to the local interfacial curvature, $\kappa$:
\begin{equation}
v = - (\lambda/2\pi) \kappa
\; ,
\label{allen-cahn}
\end{equation}
where  $\lambda$ is  a  material  constant with  the  dimensions of  a
diffusion constant, and the  factor $1/2\pi$ is for later convenience.
The Allen-Cahn equation (\ref{allen-cahn}) may be 
derived from the zero-temperature 
time-dependent Ginzburg-Landau  equation  for   
the underlying order-parameter field \cite{AC,BrayReview}. 

 
From   Eq.\  (\ref{allen-cahn}),  we   can  immediately   deduce  the
time-dependence  of  the area  contained  within  any  finite hull (i.e., 
the interior of a domain boundary) by integrating  the velocity  around 
the hull:  $dA/dt  = \oint v\,dl = -(\lambda/2\pi) \oint
\kappa\,dl = -\lambda$, the final equality following from the Gauss-Bonnet 
theorem. At  any given time  $t$, therefore, hulls with  
original enclosed-area
smaller than  $\lambda t$ will  have disappeared, and
the enclosed-areas  of surviving hulls will  have decreased by  $\lambda t$. In
other  words,  the  entire  distribution  of hull enclosed  
areas  is  advected
uniformly to the left at rate  $\lambda$. If $N_h(A,t)$ s the number of
hulls  per unit  area of  the system  with enclosed-area  greater than  $A$, it
follows that
\begin{equation}
N_h(A,t) = N_h(A+\lambda t, 0)
\; , 
\;\;\; \forall \; A>0
\; .
\label{advection}
\end{equation}

To determine  the initial  condition we note  that, shortly  after the
quench from the high temperature  phase, the system is at the critical
point of continuum percolation.  Cardy and Ziff \cite{cardy-ziff} have
shown  that the  number of  percolation hulls,  per unit  area  of the
system, with area  greater than $A$ has, for large  $A$, the universal
asymptotic form
\begin{equation}
N_p(A) \sim c/A \,
\label{cardy-ziff}
\end{equation}
where  $c =  1/8\pi\sqrt{3}$  is a  universal  constant.  This  result
provides the desired initial condition, $N_h(A,0) = 2N_p(A)$, in Eq.\ 
(\ref{advection}), giving
\begin{equation}
N_h(A,t) = 
2c/(A+\lambda t)
\; ,
\label{scaling}
\end{equation}
where the factor 2 arises from  fact that there are two types of hull,
corresponding to the two  phases, while the Cardy-Ziff result accounts
only for clusters  of occupied sites (and not  clusters of unoccupied
sites). From this result one immediately derives the hull enclosed area density 
function, $n_h(A,t) = -\partial N_h(A,t)/\partial A$, where 
$n_h(A,t)\,dA$ is the number of hulls, per unit area of the system, 
having area in the interval $(A, A+dA)$:
\begin{equation}
n_h(A,t) = 
2c/(A+\lambda t)^2
\; . 
\label{area-scaling}
\end{equation}

Equation\ (\ref{scaling}) has the expected scaling form $N_h(A,t)
= t^{-1}  f(A/t)$ corresponding to  a system with  characteristic area
proportional to  $t$. This corresponds to  characteristic length scale
$R(t)  \sim t^{\frac{1}{2}}$, which  is the  known result  if scaling  is {\em
assumed} \cite{BrayReview}.  Here, however, we do not {\em assume} 
scaling -- rather, it emerges from  the calculation.  Furthermore,  the 
conventional scaling phenomenology is restricted to  the `scaling limit': 
$A \to \infty$, $t  \to \infty$  with $A/t$ fixed. Equation~(\ref{scaling}),  
by contrast, is valid whenever $t$  is sufficiently large, and does not 
(at least on the continuum) require large $A$. This follows from the fact 
that, for large $t$, the form  (\ref{scaling}) probes, for any $A$,  the tail 
(i.e.\ the large-$A$  regime) of the Cardy-Ziff result (\ref{cardy-ziff}),  
which is  just the regime  in which the latter is valid. 

It  is, however,  instructive to  consider  what can  be deduced  from
scaling alone, augmented by  the drift equation (\ref{advection}). The
general scaling  form corresponding to a  scale area $t$  is $N_h(A,t) =
t^{-\phi}\,f(A/t)$, with arbitrary  exponent $\phi$.  Consistency with
(\ref{advection}) requires  that $N_h(A,t)$ depends on  $A$ only through
the combination $A + \lambda  t$, forcing $N_h(A,t) \propto (A + \lambda
t)^{-\phi}$.  Finally,  $\phi =  1$ is fixed  by the  requirement that
there be of order one hull per scale area, i.e.\ $N_h(0,t)
\propto 1/\lambda t$. Thus, for an internally consistent picture, one 
requires  $N_h(A,0)  \propto A^{-1}$ for large $A$,  and  it  is  gratifying 
that  the Cardy-Ziff result  not only has this form but also provides  
the exact value of the proportionality constant.

The argument above relies on the $T=0$ Allen-Cahn
Eq.~(\ref{allen-cahn}).  Temperature fluctuations have a two-fold
effect. On the one hand they generate {\it equilibrium} thermal
domains that are not related to the coarsening process. On the other
hand they roughen the domain walls thus opposing the curvature driven
growth and slowing it down.  Once equilibrium thermal fluctuations are
subtracted -- equivalently, hulls associated to the coarsening
process are correctly identified -- the full temperature dependence
should enter only through the value of the parameter $\lambda$, which
sets the time scale. For simplicity we focus here on zero working
temperature. In a future publication we shall show the
finite $T$ effects discussed above~\cite{long}. 

\begin{figure}[h]
\includegraphics[width=246pt]{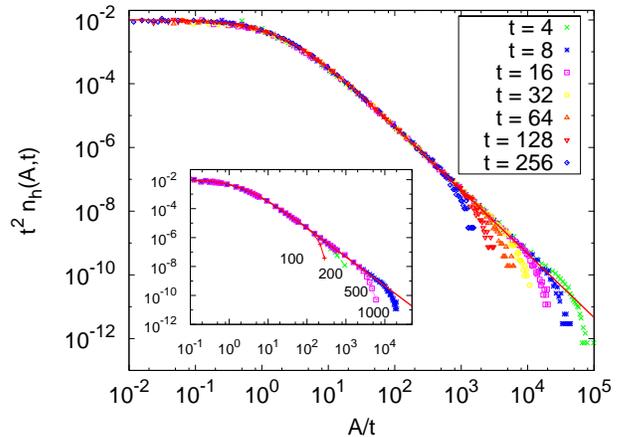}
\caption{\label{fig:hulls} (Colour online)
Number density of hulls per unit area for the 
zero temperature dynamics of the $2d$ Ising model
evolving from an infinite temperature initial condition. The full 
line is the prediction (\ref{area-scaling}) with $c=1/8\pi\sqrt{3}$ 
and $\lambda=2.1$. Inset: finite size effects at $t=16$ MCs;  
four linear sizes of the sample are used and indicated by the 
datapoints. The value of $A/t$ at which the data
separate from the master curve grows very fast with $L$ with
an exponent close to 2.}
\end{figure}

To test the above result we carried out numerical simulations on the  $2d$ square-lattice Ising model ($2d$IM) with periodic boundary
conditions using a heat-bath algorithm with random sequential updates. All data have been obtained using systems with size $L^2=10^3\times
10^3$ and $2\times 10^3$ runs using independent initial conditions.  We mimic an instantaneous quench from infinite temperature with a random
initial condition with spins pointing up or down with probability $1/2$.  The data are plotted in log-log form to test the prediction
$n_h(A,t) \propto A^{-2}$ for large $A$. The data are in remarkably good agreement with the prediction (\ref{area-scaling}) -- shown as a
continuous curve in Fig.~1 -- over the whole range of $A$ and $t$.  The downward deviations from the scaling curve are due to finite size
effects. The latter are shown in more detail in the inset where we display the $t=16$ MCs reults for four linear sizes. Finite size effects
appear only when the weight of the distribution has fallen by many orders of magnitude (7 for a system with $L=10^3$) and are thus quite
irrelevant.  The only fitting parameter is $\lambda$, which has the value $\lambda=2.1$ in Fig.~1.  The agreement between theory and data is
all the more impressive given that the curvature-driven growth underlying the prediction (\ref{area-scaling}) only holds in a statistical
sense for the lattice Ising model \cite{Anisotropy}.

The mean hull enclosed  area, per unit area of the system, diverges
with the system size: $\langle A \rangle = \int_0^{L^2}
dA\,A\,n_h(A,t) \sim 2c \ln (L^2/\lambda t)$.
The fact that the total hull enclosed area exceeds the area of the
system seems paradoxical until one notes that a given point in space
belongs to many hulls.

It is clear that the evolution of the hull-enclosed area distribution follows
the same `advection law' (\ref{advection}), with the same value of
$\lambda$, for other initial conditions.
Moreover, Eq.~(\ref{area-scaling}) applies 
to {\it any} $T_0>T_c$ equilibrium initial condition asymptotically. Equilibrium initial 
conditions at different $T_0>T_c$ show only a different transient behaviour; 
initial states that are closer to $T_c$ take longer to reach the 
asymptotic law~(\ref{advection}) (see the inset in Fig.~2)~\cite{long}.  

An equilibrium state at the critical temperature,
$T_0=T_c$, is, as expected, different.  This case has already been
addressed, for general space dimension, in the context of coarsening
from an initial state with long-ranged spatial correlations
\cite{bray-humayun-newman}.  It was argued that the characteristic
scale $R(t)$ still grows as $t^{1/2}$, since the growth is curvature
driven, but the space and time correlation functions are modified if
the spatial correlations in the initial state are sufficiently
long-ranged \cite{bray-humayun-newman}.  The statistics of hull and
domain areas, however, were not discussed.  In place of the continuum
percolation statistics that characterize the initial state {\it shortly}
after a
quench from high temperature, the initial state for a quench from
$T_c$ is characterized by the statistics of Ising cluster hulls at the
critical point.  The area distribution of the hulls of these clusters
has been studied by Cardy and Ziff \cite{cardy-ziff}.  The result has
a form identical to Eq.\ (\ref{cardy-ziff}), but with $c$ replaced by
$c/2$. We predict, therefore, that our results for the hull-enclosed 
area distributions can be generalized to this case by simply making
the replacement $c \to c/2$ everywhere, and keeping the value of
$\lambda$ unchanged.  In Fig.~\ref{fig:three} we compare this
prediction to the number density of hull-enclosed areas in the
$2d$IM evolving at $T=0$ from a critical initial condition,
once more obtaining excellent agreement.

\begin{figure}[h]
\includegraphics[width=246pt]{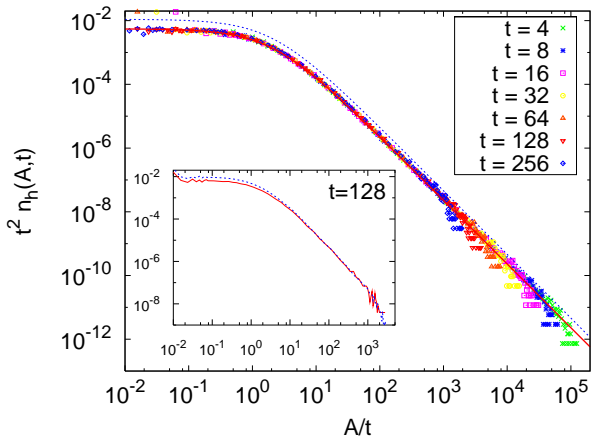}
\caption{\label{fig:three} (Colour online) Number density of hulls 
per unit area for the zero-temperature $2d$IM
 evolving from critical initial conditions. We obtained the
initial states after running $10^3$ Swendsen-Wang algorithm  steps. The
full (red) line represents (\ref{area-scaling}) with $c \to c/2$ and 
$\lambda = 2.1$. The dashed (blue) line is (\ref{area-scaling}) with $c$.
Inset: comparison between the hull-enclosed area distribution 
at $t=128$ MCs for equilibrated initial conditions at $T_0=2.5$ and 
$T_0\to\infty$.}
\end{figure}

Also interesting is the distribution of {\em domain} areas,
$n_d(A,t)$, which are the areas of regions of aligned
spins~\cite{Liverpool}. Domains are obtained from hulls by removing
any interior hulls.  The domain area distribution $n_d(A,t)$ (number
density of domains with area $A$, per unit area) of the $2d$IM
evolving at zero temperature after a quench from infinite
temperature is shown in Fig.~3. The downward deviations from the
scaling curve in the main panel as well as the bumps on the tail are
finite size effects due to the percolating clusters and, as 
for the hull-enclosed areas (see Fig.~1), are not very important.

\begin{figure}[h]
\includegraphics[width=246pt]{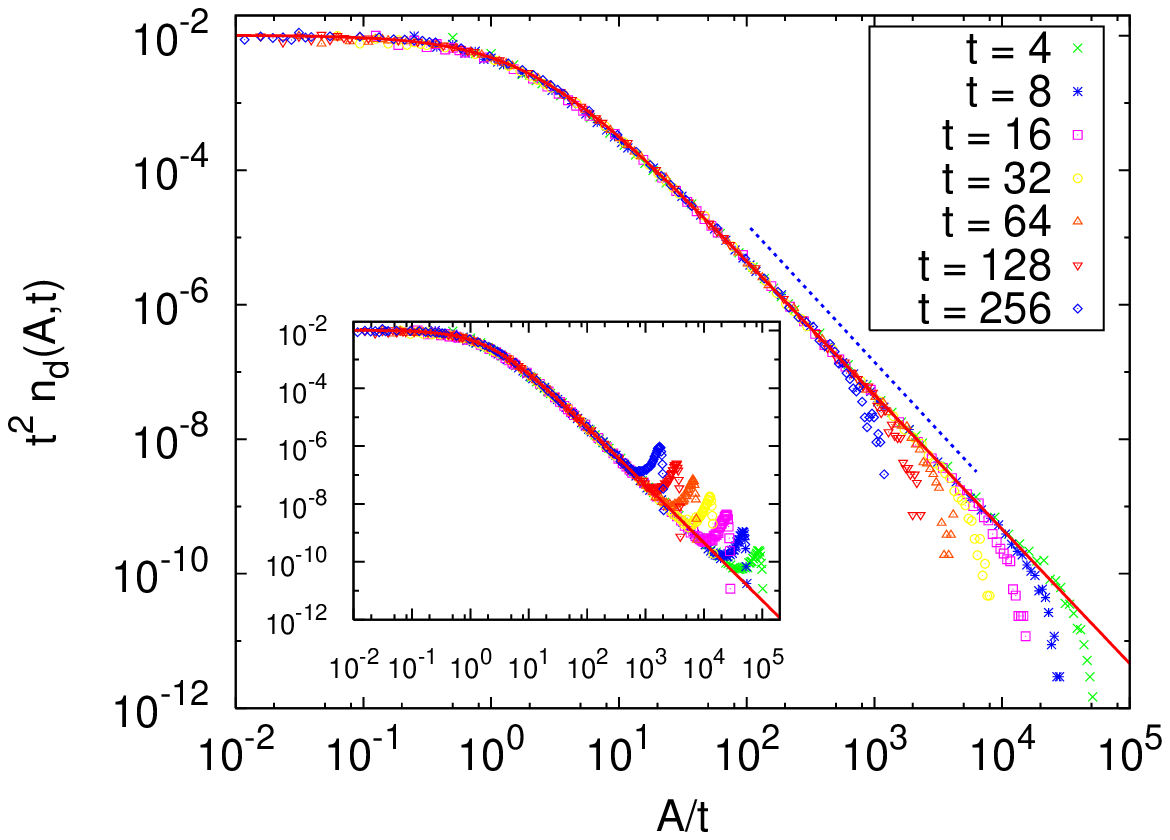}
\includegraphics[width=246pt]{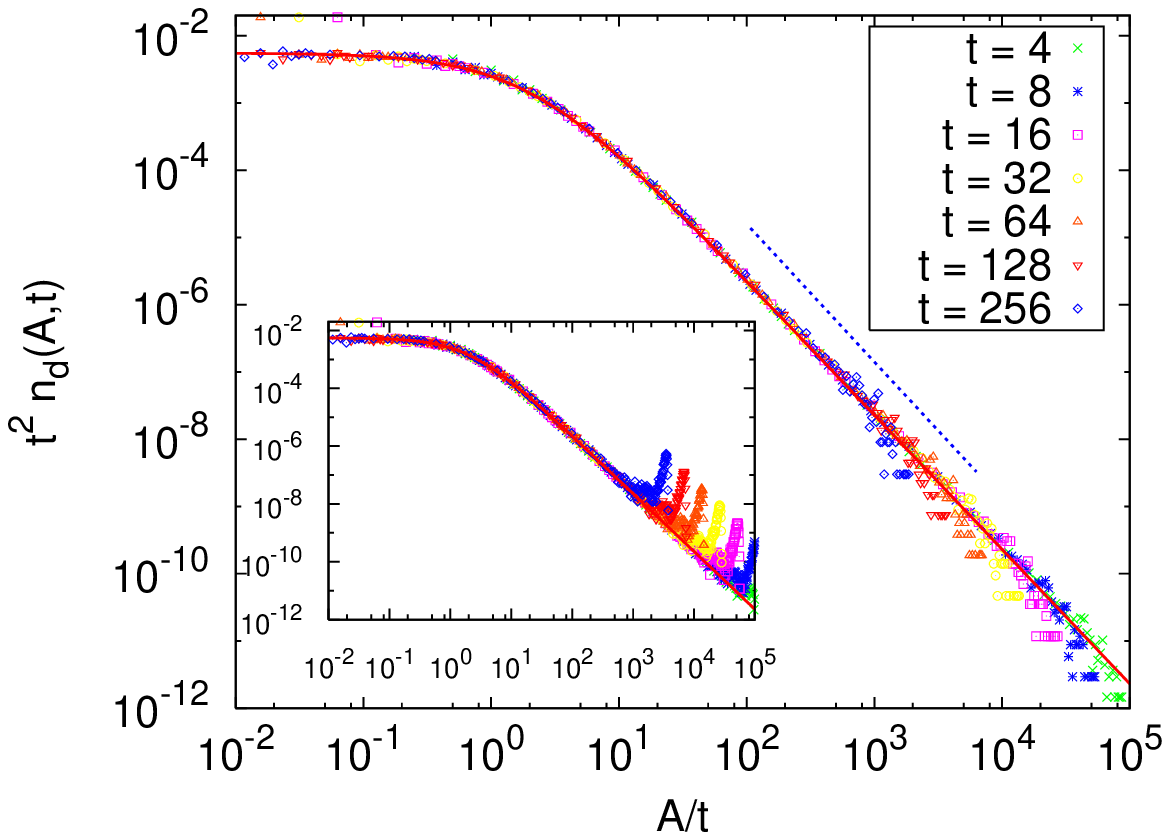}
\caption{\label{fig:three} \label{fig:domains} 
(Colour online) Number density of 
domains per unit area for the zero-temperature $2d$IM
evolving from $T_0\to\infty$ (upper panel) and $T_0=T_c$ (lower panel) initial conditions. 
In the main panels the percolating domains have been extracted from 
the analysis while in the insets we show the same data including the 
percolating clusters.
We obtained the
initial states after running $10^3$ Swendsen-Wang algorithm  steps. The
full (red) line represents (\ref{area-scaling}) 
with $c=1/8\pi\sqrt{3}$ (upper) and $c\to c/2$ (lower),  
and $\lambda=2.1$ in both cases. 
The dotted (blue) lines have slopes -2.03 (a) and -2.05 (b).}
\end{figure}

Remarkably, the domain area distribution, $n_d$, seems to be almost identical to the hull-enclosed area
distribution $n_h$, {\it i.e.}  $n_d \sim 2c_d/(A+\lambda_d t)^\tau$ with a prefactor $2c_d$, a parameter
$\lambda_d$ and an exponent $\tau$ taking approximately the same values as for the hull-enclosed  areas.
An argument that treats interior domain walls in a mean-field approximation and uses the 
exact result for $n_h$ derived above, allows one to
derive~\cite{long} 
\begin{equation} 
n_d(A,t)\sim 2c_d [\lambda_d (t+t_0)]^{\tau-2}/[A+\lambda_d (t+t_0)]^\tau \; ,
\end{equation} 
with $\lambda_d = \lambda +O(c)$, $c_d=c+O(c^2)$ and $\tau\sim 2+O(c)$ for infinite
temperature initial conditions and, $c_d \to c_d/2$ for initial conditions equilibrated at $T_0=T_c$. 
$t_0$ is such that $\lambda_d t_0 =a^2$, with $a$ a microscopic length-scale, 
and sets the time scale.
The constant $c_d$ and $\tau$ 
characterize the initial distribution of domain areas,
$n_d(A,0) \sim 2c_d a^{\tau-2} A^{-\tau}$. The exponent $\tau$ is known
for the $2d$IM critical geometrical clusters, $\tau= 379/187\approx 2.03$~\cite{Stella}, and for cluster
masses in $2d$ percolation, $\tau = 187/91 \approx 2.05$~\cite{Stauffer}; the constant $c_d$ has not
been computed analytically.  $\tau>2$ allows one to satisfy the sum rule establishing that the total
domain area, per unit area of the system, must be equal to unity since every spin belongs to one, and only
one, domain.  Unfortunately, it is hard to distinguish a power law with $\tau=2$ or 2.03-2.05 from
our numerical data: both exponents describe the power-law tail equally well, as shown in
Fig.~\ref{fig:domains}, with a (blue) dotted line.  It is, however, possible to put to the numerical test
the value of the prefactor $c_d$ (or $2c_d$) and $\lambda_d$. 
We shall present this analysis elsewhere~\cite{long}.

\begin{figure}[h]
 \includegraphics[width=246pt]{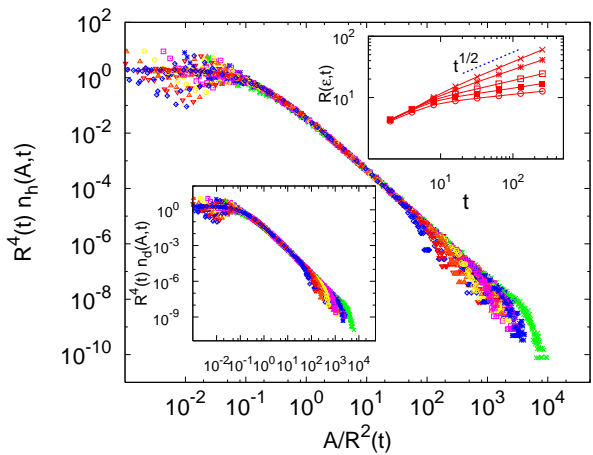}
\caption{\label{fig:four} 
(Colour online) Number density of hull-enclosed and domain 
(lower inset) areas per unit area for the $2d$ random
ferromagnetic Ising model with a uniform distribution of exchanges in 
the interval $[2-\epsilon, 2+\epsilon]$ with $\epsilon=0,\,0.5,\,1,\,1.5,\,2$. 
The system evolves at $T=0.4$ after a
quench from $T_0\to\infty$. Upper inset: the evolution of 
$R(\epsilon,t)$ used 
to scale the pdfs and its comparison to the pure $R(0,t) \propto t^{1/2}$ law.} 
\end{figure}

Interestingly, similar results are obtained for the $2d$ {\em random}
ferromagnet, at least when activation is not too
important. In Fig.~\ref{fig:four} we display data obtained at $T=0.4$
after a quench from $T_0\to\infty$. This low but finite working
temperature is enough to avoid the complete pinning of domain walls
by quenched disorder. Good scaling with the typical domain area $R^2(t)$
is obtained for $A/R^2(t) \geq 10^{-1}$ and the master curve resembles strongly 
the one found in the pure Ising case also included in Fig.~\ref{fig:four}. 

In summary, we have obtained exact results for the statistics of
 hull-enclosed areas during the phase-ordering dynamics of $2d$
systems coarsening under curvature driven growth. Notably, these
results include a proof of the scaling hypothesis for these systems.
The results are strongly supported by simulations of Ising systems,
suggesting a strong degree of universality.
The domain area distribution satifies a very similar law.

AJB and LFC thank R. Blumenfeld, J. Cardy, D. Dhar, C.
Godr\`eche, S. Majumdar, P. Sollich, and R. Stinchcombe for
very useful discussions. AJB and LFC thank the Isaac Newton Institute,
Cambridge, and JJA the LPTHE Jussieu, Paris, for their hospitality
during the period when this work was being carried out. JJA, LFC and
AS acknowledge financial support from Capes-Cofecub research grant
448/04. JJA is partially supported by 
CNPq and FAPERGS and LFC by IUF.


\end{document}